# The Concentration Limit for Solar Cells Based on Entropy Production

Ze'ev R. Abrams and Xiang Zhang

*Abstract*— The maximal concentration limit of solar radiation for photovoltaic applications is assumed to be constant, at $C^{max} \approx 46,000$ suns. This limit is easily found via a geometrical application of the 2$^{nd}$ law of thermodynamics to a radiation transfer system [A. Rabl, *Sol. Energy* vol. 18, pp. 93-111, 1976]. However, previous analysis did not include the generation of entropy in photovoltaic conversion. Here, we show a bandgap dependence of the maximal concentration limit for a semiconductor solar cell when taking this entropy generation into account, and show that the limit is reduced for low bandgap materials. This new concentration limit lies in contrast to the assumed invariance of the concentration limit, and we attribute this difference to a breakdown in the assumptions used to derive the traditional detailed-balance model.

## I. Introduction

Increasing the production of solar energy will require the use of solar concentrator systems that better utilize the solar radiance [1]-[3]. Concentrator systems mimic the effect of adding more suns to the sky, such that the solar cell effectively absorbs more solar photons per unit area. The limitation to this concentration has been known to lie in a geometric application of the 2$^{nd}$ law of thermodynamics [3], which has been associated with the brightness theorem [4], or the concept of optical étendue [5]. This limit is assumed to be invariant, regardless of the material system used for the cell. In this work, we show that this assumption does not hold for photovoltaics when the production of entropy is included into the detailed-balance model. We show that this variance is a result of incorrect assumptions used to derive the maximal voltage efficiency of a solar cell.

We first review the analysis providing the original limit of concentration [3], [6]. The concentration, $C$, is a ratio of the aperture of the concentrator, $A_{conc}$ to that of the surface aperture of the solar cell, $A_{cell}$:

$$C = A_{conc} / A_{cell} \qquad (1)$$

The amount of radiation from the sun, at temperature $T_S$=6000 K, and reaching the front aperture of the concentrator system is related to the distance between the sun and the earth, $R$, and the radius of the sun, $r$ [6]:

$$Q_{conc} = \frac{4\pi r^2}{4\pi R^2} A_{conc} \sigma T_S^4 \qquad (2)$$

with $\sigma$ being the Stefan-Boltzmann constant, and the ratio of areas relating the reduction of radiation flux from the solar surface to the surface of the concentrator on earth. The solar disc has an angular radius of $\Delta_S \approx 4.7$ mrad, such that $\sin\Delta_S = r/R$. Furthermore, the incoming étendue of the solar radiation has a solid angle of $\Omega_S = \pi \times \sin^2\Delta_S \approx 6.85 \times 10^{-5}$ sr. In contrast, the cell at temperature $T_c$ and index of refraction $n_c$ will radiate given by [6]:

$$Q_{cell} = A_{cell} n_c^2 \sigma T_c^4 \qquad (3)$$

The outgoing étendue emission from the cell is $\Omega_o$ and can be taken as $\pi n_c^2$ for a flat plate cell with a back reflector, or $2\pi n_c^2$ for a bifacial cell.

The 2$^{nd}$ law of thermodynamics can be applied using Clausius' statement that (in the absence of external work), the flow of heat must be from hot to cold bodies. In mathematical form, using the concept of entropy, $S$, this is stated as:

$$\Delta S = \Delta Q/T \geq 0 \qquad (4)$$

Given that the flow of heat is from the concentrator to the cell, this defines $\Delta Q \equiv Q_{conc} - Q_{cell}$. Clearly the minimum possible entropy generation is found for the case of equal temperatures, such that $Q_{conc} - Q_{cell} = 0$ when $T_S = T_c \equiv T$. This minimal entropy case occurs when (2) is equal to (3) [assuming no other temperature source, such as the ambient blackbody from the earth is present], resulting in:

$$\sin^2\Delta_S A_{conc} \sigma T^4 = A_{cell} n_c^2 \sigma T^4 \qquad (5)$$

Manuscript submitted April 25$^{th}$ 2012. This work was supported by the U.S. Department of Energy, Basic Energy Sciences Energy Frontier Research Center (DoE-LMI-EFRC) under award DOE DE-AC02-05CH11231. ZRA acknowledges Government support under and awarded by DoD, Air Force Office of Scientific Research, National Defense Science and Engineering Graduate (NDSEG) Fellowship, 32 CFR 168a.

Z.R.A. and X.Z. are with the Applied Science & Technology program at the University of California, Berkeley USA and the Materials Science Division, Lawrence Berkeley National Laboratory, 1 Cyclotron Road, Berkeley, California 94720, USA

Current address for ZRA (corresponding author): zabrams@stridersolar.net.

Inserting (5) into (1) results in the maximal concentration limit:

$$C^{max} = n_c^2 / \sin^2 \Delta_S = \Omega_o / \Omega_S \qquad (6)$$

where a factor of 2 can be included to distinguish the reflector and bifacial systems. For most cases, it is assumed that the index of refraction is unity, such that $C^{max} \approx 46,000$ (rounding up) for a flat plate cell with a single face/back reflector. This concentration limit is assumed to be *invariant*, and is seen to be a result of a single application of the 2nd law of thermodynamics to the most basic of heat transfer systems involving emission and absorption. This concentration limit can also be described using brightness theory to claim that the light cannot be bent backwards, such that the change in étendue cannot exceed a 90° angle [4].

When $T_c=T_S$, no finite power can be extracted from the cell, since there is no free energy left to be converted in the system. However, in a solar conversion system, the actual temperature of the cell will be related to the amount of concentration by assuming that some of the incoming radiation is converted to other forms of energy (electrical, chemical, heat, etc), at an extraction efficiency $\eta_{ext}$ (this is *not* the efficiency of the solar cell). Under this basic assumption, the energy balance is achieved by equating (2) with (3) in addition to a fraction of the heat extracted, $\eta_{ext}Q_{conc}$:

$$Q_{conc} = Q_{cell} + \eta_{ext}Q_{conc} \qquad (7)$$

In this equation, it is assumed that Kirchhoff's law of radiation applies, such that the absorptivity is equal to the emissivity, $\alpha_{abs}=\varepsilon_{emit}$. Inserting (2) and (3) into (7) relates the temperature of the cell to the concentration [3], [6]:

$$T_c \cong T_S \sqrt[4]{[1-\eta_{ext}]C/C^{max}} \qquad (8)$$

The temperature of the cell will only reach the temperature of the sun at maximal concentration, and with no other extraction losses. However, no photovoltaic energy conversion will be obtained at this temperature.

## II. ENTROPY PRODUCTION USING DETAILED-BALANCE

The efficiency of a solar cell is found using the detailed-balance model [7], which equates the incoming and outgoing radiation flux for the cell using a two-band model of a semiconductor [8], [9]. The incoming radiation flux, $N_{in}$, is a function of the blackbody radiation from the sun, as well as the contribution from the ambient blackbody, at $T_o=300$ K. This is given by Planck's blackbody formulas as:

$$N_{in}(C) \propto C\Omega_S \int_{E_g}^{\infty} \frac{E^2 dE}{\exp[E/kT_S]-1} + (1-f_\Omega C)\Omega_o \int_{E_g}^{\infty} \frac{E^2 dE}{\exp[E/kT_o]-1} \qquad (9)$$

In (9), $k$ is Boltzmann's constant and $E_g$ is the bandgap of the semiconductor. The incoming flux is only taken from the bandgap, assuming a step-function dependence of the absorption coefficient, $\alpha_{abs}$, on the bandgap, $E_g$ [7], [9]. The ratio, $f_\Omega=\Omega_S/\Omega_o$ is used to include only the portion of the ambient blackbody that lies outside the beam of incoming light. The constant of proportionality in (9) is $2/h^3c^2$, with $h$ being Planck's constant, and $c$ being the speed of light in vacuum. The input radiation flux can also be taken from existing tables for the actual spectrum incident on the cell, such as the AM 1.5 spectrum.

The outgoing flux emission is proportionate to the van Roosbroeck-Shockley relation for a semiconductor [10], [11], and includes the chemical potential of the cell:

$$N_{out} \propto \Omega_o \int_{E_g}^{\infty} \frac{E^2 dE}{\exp[(E-\mu)/kT_c]-1} \qquad (10)$$

Here it is assumed that $n_c=1$, and that the outgoing emissivity is assumed to be equal to the absorptivity ($\alpha_{abs}=\varepsilon_{emit}$), which provides the lower boundary to the integral in (10). Equating the two photon fluxes at open-circuit [9], when no current is extracted, and using a well-known approximation of the integral when $E-\mu>>kT_c$, one can obtain a closed-form equation for the chemical potential [12]:

$$\mu_{oc} = E_g - kT_c \cdot \ln\left[\frac{\Omega_o E_g^2 kT_c \alpha_{c1}}{N_{in}(C)}\right] \qquad (11)$$

Note that this approximation neglected the "-1" term in the denominator. While this approximation is well known, and has been used in many other previous results [5], [9], [16], the approximation's value begins to break down at low bandgaps (when $E-\mu \sim kT_c$).

The incoming flux has been retained in its integral form, $N_{in}(C)$ from (9), and $\alpha_{c1}=1+2kT_c/E_g+2(kT_c/E_g)^2$ is a correction term stemming from a more accurate approximation of the integral [5], [13]; $\alpha_{c1} \approx 1$ for values of bandgap that are much larger than the thermal energy, which is $\approx 25.8$ meV for temperatures of $T_c=300$ K. The chemical potential at open-circuit is related to the open-circuit voltage by $V_{oc}=\mu_{oc}/q$, with $q$ being the electric charge constant.

The chemical potential at open-circuit relates the chemical potential of the electrons in the cell to the chemical potential of the photons after being emitted from the semiconductor [5], [9]. The chemical potential of the photons need not be zero after their interaction with matter [14]. At open-circuit, the free energy available for photovoltaic conversion is maximal however the efficiency of conversion will be zero, since no current can be extracted and $P=I\times V$. This chemical potential can be related to the Gibbs Free Energy of the system [15], [16], $G$, by the fundamental relation:

$$G = U - T_c S = \mu_{oc} \qquad (12)$$

where $U$ is the internal energy, and $S$ is the entropy. Comparing (11) and (12), one can recognize the entropic contribution to the chemical potential as being the term $k\times ln[\ ]$. However, the more precise derivation of the entropy is found by taking the negative partial derivate with respect to the temperature: $S=-\partial G/\partial T_c$, (see Appendix for comparison) which results in:

$$S = k\frac{\alpha_{c2}}{\alpha_{c1}} + k\ln\left[\frac{\Omega_o E_g^2 k T_c \alpha_{c1}}{N_{in}(C)}\right] \qquad (13)$$

where $\alpha_{c2}=1+4kT_c/E_g+6(kT_c/E_g)^2$ is a result of the derivative of the term $\partial(T_c\alpha_{c1})/\partial T_c$, and $N_{in}(C)$ is assumed to not be dependent on $T_c$ under the assumption that $T_o \neq T_c$. The correction terms $\alpha_{c1}$ and $\alpha_{c2}$ are typically taken as unity, however are not done so here, where small bandgaps are considered as well.

The 2nd law of thermodynamics, in its most general form, states that the production of entropy in a system must increase: $\Delta S \geq 0$. Therefore, we can apply the 2nd law of thermodynamics a *second* time, and enforce $S\geq 0$ to obtain the following relation from (13):

$$e^{\alpha_{c2}/\alpha_{c1}} \Omega_o E_g^2 k T_c \alpha_{c1} \geq N_{in}(C) \qquad (14)$$

Equation (14) is thus a rephrasing of the 2nd law of thermodynamics, in the context of broadband solar illumination on a semiconductor with concentration taken into account. Since the constants in the equation are determined externally (such as the temperatures involved), the value of concentration, $C$, can create a condition for which the entropy in the system is negative. This violation of the 2nd law of thermodynamics is a function of the *bandgap* and *temperature* of the cell, and a maximal concentration factor per bandgap can be found by locating the point at which the inequality in (14) is reversed. This is plotted in Fig. 1, for values of $\Omega_o$ of $2\pi$ (solid line) and $\pi$ (dotted line).

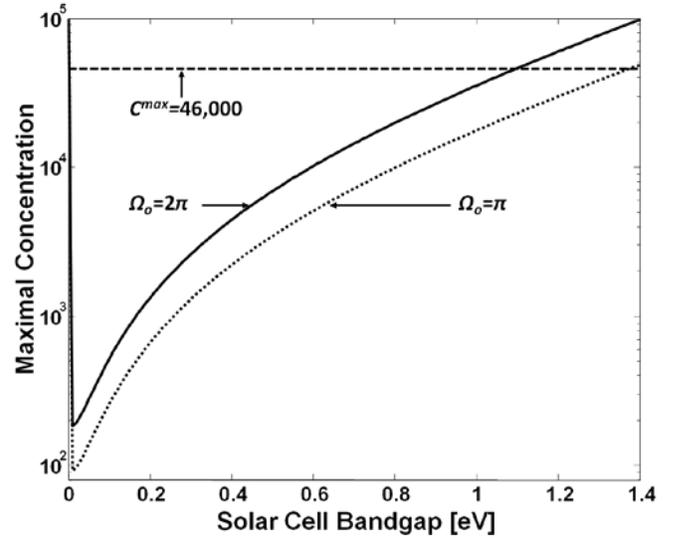

Fig. 1. Maximal concentration per bandgap energy. The maximal concentration is plotted for two geometries of cells: a flat plate ($2\pi$ emission, solid line), and a flat plate with a reflector ($\pi$ emission, dotted line). The upper limit of 46,000 generally given as the maximal concentration is plotted as the horizontal dashed line. The temperature of the cell here is taken as Tc=To=300 K. At bandgaps approaching the thermal energy of 25.8 meV, the equation breaks down, as is seen at the far left.

The maximal concentration of 46,000 (or $\pi/\Omega_{Sun}$) shown in Fig. 1 (dashed horizontal line) is no longer be an invariant constant, rather is dependent upon the bandgap of the material in conjunction with the 2nd law, as expressed in (14). The apparent violation of the 2nd law is counter intuitive, since there is seemingly no limitation to placing a low bandgap semiconductor within a system with very high concentration. While it is known that a practical limit of $C^{max} \approx 1000$ suns appears due to parasitic losses in the series resistance in most cells [2], [17], the violation of the 2nd law appearing here in Fig. 1 is based only on the most basic assumptions governing the detailed-balance formalism.

III. PHYSICAL INTERPRETATION OF THE NEW CONCENTRATION LIMIT

A closed-form formulation of (13) can provide a physical explanation for the origin of this new concentration limit. If we remove the contribution of the ambient blackbody from the input flux of (9), which provides a negligible contribution of flux under high concentration [18], and use the approximation of the integral as above, assuming that $E>>kT_S$, we can obtain a well-known formula for the open-circuit chemical potential of the cell [5], [9], [13], [16]:

$$\begin{aligned}\mu_{oc} &= E_g(1-T_c/T_S) \\ &+ kT_c \cdot \ln\left[(C\Omega_S/\Omega_o)(T_S/T_c)(\alpha_{S_1}/\alpha_{c1})\right]\end{aligned} \qquad (15)$$

here, $\alpha_{S1}=1+2kT_S/E_g+2(kT_S/E_g)^2$ [13]. The term $(1-T_c/T_S)$ is the Carnot efficiency, which is considered as the maximal obtainable voltage efficiency from a solar cell [19]. From (15), if we neglect the correction terms ($\alpha_1$ and $\alpha_{c1}$), the entropy is seen to primarily consist of a negative contribution due to the reduction in étendue ($C\Omega_S/\Omega_{out}$), and a positive contribution due to the reduction in temperature of the photons from $T_S$ to $T_c$ [5], [16]. Under maximal concentration conditions, the entropy term cancels the effect of the Carnot efficiency term, resulting in $qV_{oc} \to E_g$, particularly when $T_c \to 0$ K, which is the "Ultimate Efficiency" limit of a solar cell [7].

Under the approximations used for (15), we can again derive the entropy by taking the partial derivative with respect to the temperature and enforce $\Delta S \geq 0$, resulting in:

$$\frac{E_g}{kT_S} + \frac{\alpha_{c2}}{\alpha_{c1}} + \ln\left[\frac{\Omega_o T_c \alpha_{c1}}{\Omega_S T_S \alpha_{S1}}\right] - \ln[C] \geq 0 \qquad (16)$$

This results in a closed-form formula for the maximal concentration, as a function of the relevant parameters of the system:

$$C^{max} = \frac{\Omega_o T_c}{\Omega_S T_S}\left(\frac{\alpha_{c1}}{\alpha_{S1}}\right)\exp\left(\frac{E_g}{kT_S}\frac{\alpha_{c2}}{\alpha_{c1}}\right) \qquad (17)$$

The formula given in (17) holds for $E_g > kT_S = 0.517$ eV, however is numerically similar to (14) even for values of $E_g$ approaching zero. We thus obtain a simple relation between the maximal concentration and the bandgap, incoming and outgoing étendues, and temperatures. Since $\alpha_{c1} \approx \alpha_{c2} \approx 1$, for most $E_g > kT_c$, the contributions of the $\alpha$ terms are mostly negligible.

The closed-form result of (17) is in contrast with the traditionally held form of the maximal concentration given in (6) above, which stated that $C^{max} = \Omega_o/\Omega_S$. However, to derive (6), it was assumed that the temperature of the cell was equal to that of the sun ($T_S = T_c$). Using this assumption in (17), and neglecting the term $\alpha_{c2}/\alpha_{c1}$, we obtain:

$$C^{max}|_{T_S=T_c} = \frac{\Omega_o}{\Omega_S}\exp\left(\frac{E_g}{kT_S}\right) \\ = C_0^{max}\exp\left(\frac{E_g}{kT_S}\right) \qquad (18)$$

Equation (18) relates the maximal concentration of a cell consisting of a semiconductor with bandgap $E_g$, to that of a regular material, which can absorb all the heat/light emitted from the sun. The "original" concentration, $C_0^{max} = \Omega_o/\Omega_S$, is relevant only for a material that can absorb *all* the incoming heat; however a semiconductor will only absorb the photons from above the bandgap, thereby changing the maximal concentration amount, even if the temperatures are equal. For a metal, with $E_g = 0$, absorbing all the incoming radiation, we obtain from (18) the traditional limit. The concentration limit is therefore a function of the spectrum itself, as we have recently shown for the case of down-conversion [20].

## IV. MODIFYING THE DETAILED-BALANCE ASSUMPTIONS

The result of Fig. 1 lie in contrast with physical intuition: there is no physical limitation to concentrating the light at $C_0^{max}$ at any point in space, resulting in a localized temperature of $T_S$. However, in the basic assumptions used to derive the detailed-balance model for a solar cell, the implicit assumption was made that the cell is maintained at the ambient temperature, $T_c = T_o$, and did not include any relation between the internal temperature of the cell with the concentration factor, as presented in (8). Therefore, a paradox is created when the cell is placed at the focus of the maximally concentrated solar illumination, while being forced to radiate outward using (10), and maintained at $T_c = 300$ K.

The entropic paradox can be mitigated when allowing for the temperature of the cell to rise. Since we do not know *a priori* the amount of heat that will be extracted in (8), we can leave the temperature of the cell, $T_c$, variable when calculating the concentration limit. Plotted in Fig. 2 is the maximal concentration using (14) when allowing the temperature of the cell to rise. As is shown, the entropic restriction no longer occurs at cell higher temperatures.

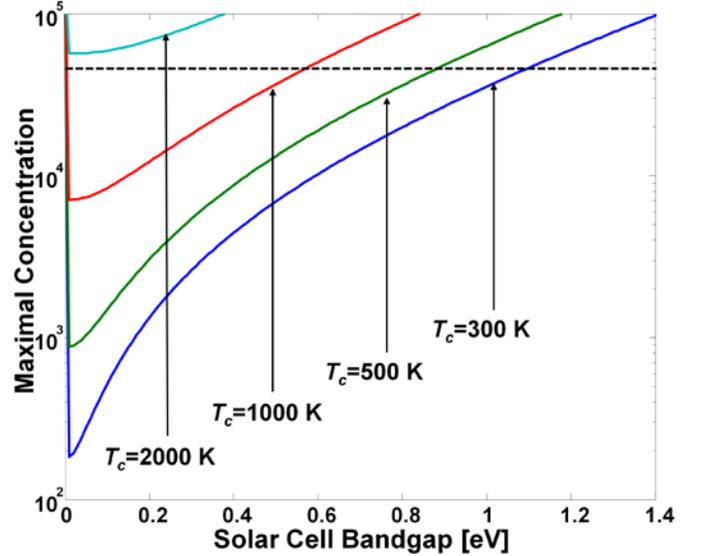

Fig. 2. (Color Online) Maximal concentration per bandgap energy, for rising temperatures of the cell. The graph is plotted similar to Fig. 1, with only the $\Omega_o = 2\pi$ contribution depicted.

The temperature rise within the semiconductor will be a result of the absorption of photons from the band-edge and up, with the excess energy transferred to phonons in the lattice in a thermalization process. The flux equilibrium method [9], as well as the original detailed balance analysis [7], uses a model with continuous conduction and valence bands [8], but does not consider this thermalization loss process. In addition to thermalization, free carrier absorption within these bands will reduce the number of photons contributing to the utilizable power, since the intraband absorption processes create phonons instead of electron-hole pairs, reducing the utilizable electrons as occurring in other low bandgap devices [21]. Free-carrier absorption will also generate heat due to low energy photons (with energy below the bandgap, typically in the infra-red regime), which are absorbed in intraband transitions. These heat generation mechanisms will force the cell's lattice temperature to rise, particularly for low bandgap devices, and will invalidate the assumption that the cell's temperature is maintained at 300 K.

In addition to the direct production of heat, which will challenge the original assumptions used for the detailed-balance model, other mechanisms will prevent this entropy violation from being seen experimentally. For example, the bandgap of a semiconductor material is typically strongly a function of the temperature. For temperatures of the cell approaching solar temperatures, the bandgap will drop significantly, modifying the assumptions used above. Low bandgap materials also have a high index of refraction, and therefore higher reflectivity losses, negating the perfect absorption assumption. Furthermore, the inclusion of the index of refraction, which incurs an additional concentration factor in the entropy of the order of $4n_c^2$ [23], was not considered here.

The approximation of the integral such that $E-\mu \gg kT_c$ will not hold under high concentration levels, where the chemical potential rises and approaches the bandgap of the material, $\mu \rightarrow E_g$ [19], [22]. For high concentration levels and low bandgaps, the difference $E-\mu$ is on the order of the thermal energy, and degenerate carrier concentrations are induced; (10) then approaches a one-sided delta function. At this point, the assumptions that were used for the detailed-balance model between the rate of stimulated absorption, $r_{abs}$, and spontaneous emission $r_{spon}$, are no longer valid, since we must include the rate of stimulated emission, $r_{stim}$, [7], [24], and write: $r_{abs}=r_{spon}+r_{stim}$ to balance the photon fluxes. The inclusion of the stimulated emission is particularly important for very low bandgap semiconductor ($E_g<0.5$ eV) detectors, where the spectra of the thermal and radiative emission overlap [21].

## V. Conclusions

The thermodynamic limits of concentration have also been found for monochromatic fluorescent systems [25], [26]. In particular, second applications of the $2^{nd}$ law of thermodynamics to limit the maximal concentration for these systems have been applied, such that the maximal Stokes shift is concentration limited. These formulas relied on a monochromatic absorption spectrum; however, applying a concentration limitation to the *broadband* solar spectrum as well as to a regular, high bandwidth solar cell has not been done before, such that the concentration is shown to be bandgap dependent.

The maximal concentration for most feasible solar cell designs is still well above that of any pragmatic design, once losses are considered. Specifically, series resistance losses circumvent most solar concentration above 1000 suns to be used before seeing a dramatic decrease in power conversion efficiency [2], [17]. However, the limitation on high concentration for low bandgap materials raises the question whether many recent ideas of employing novel, low bandgap materials such as carbon nanotubes or graphene for solar cells [27], [28] is possible without requiring significant external cooling mechanisms. Furthermore, the high efficiency values reported in many theoretical works, approaching the Carnot efficiency limit [19], may be further limited by this *second* application of the $2^{nd}$ law of thermodynamics, thereby reducing the maximal power utilizable from the sun for future needs. This would apply to other $3^{rd}$ generation photovoltaic concepts [29] as well, where maximal efficiency is found for very low bandgap devices. Verifying that the generation of entropy is not violated should be confirmed for all manners of photovoltaic conversion [30], particularly for methods requiring high concentration, without providing excessive external energy for cooling the system.

One can therefore view the results of this work as either a result of basic thermodynamic arguments, with (18) being a new variation of the maximal concentration, or as a paradox resulting from the assumptions of a room-temperature cell concentrated to the temperature of the sun, while only including spontaneous emission. The approximation of the integral (neglecting the "-1" term) still applies for larger bandgaps, where (18) still shows some effect.

## Appendix

Associating the entropy directly with the $k \times ln[]$ term in (13) will result in a truncated and approximated version of the entropy:

$$S' = k \ln\left[\frac{\Omega_o E_g^2 kT_c \alpha_{c1}}{N_{in}(C)}\right] \qquad (19)$$

This equation for the entropy is missing the additional contribution of $k \times \alpha_{c2}/\alpha_{c1} \approx k$. If we apply the $2^{nd}$ law of thermodynamics to (19), such that $\Delta S \geq 0$, we will require that the fraction in the brackets be larger than unity, which is equivalent to (14), without the term $exp(\alpha_{c2}/\alpha_{c1})$. This is plotted in Fig. 3, for $T_c=300°$ K and $\Omega_o=2\pi$. The result is quite similar to that of Fig. 1, with a more stringent limitation on the maximal concentration in the lower bandgap regime.

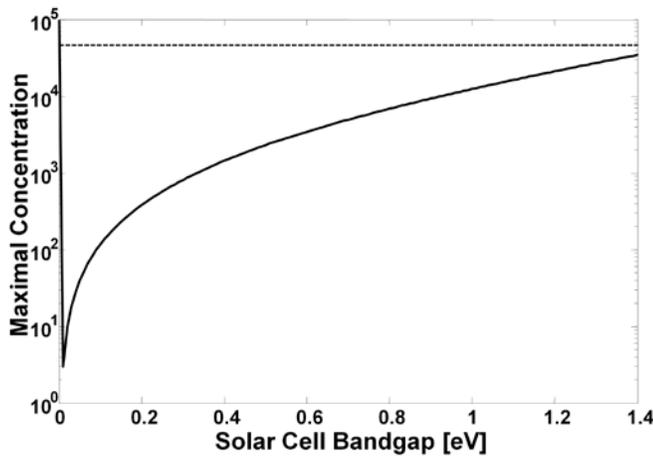

Fig. 3. Maximal concentration using the approximation for the entropy at 300° K and $\Omega_o=2\pi$. This graph has a lower limit of maximal concentration for small bandgaps, in contrast with Fig. 1.

ACKNOWLEDGMENT

Z.R.A. would like to thank T. Markvart J. Gordon, O. Miller and M. Sheldon for their critical comments.